\documentclass[twocolumn,prx,floats,superscriptaddress]{revtex4-1}
\newcommand{\tw}{\ensuremath{t_\mathrm{w}}\xspace}
\newcommand{\twone}{\ensuremath{t_\mathrm{w1}}\xspace}
\newcommand{\twtwo}{\ensuremath{t_\mathrm{w2}}\xspace}

\newcommand{\Tg}{\ensuremath{T_\mathrm{g}}\xspace}

\newcommand{\teff}{\ensuremath{t^\mathrm{eff}_\mathrm{w}}\xspace}
\newcommand{\teffw}{\ensuremath{t^\mathrm{eff}_\mathrm{w}}\xspace}

\newcommand{\xiZ}{\ensuremath{\xi_\mathrm{Zeeman}}\xspace}
\newcommand{\ximN}{\ensuremath{\xi_\mathrm{micro}^\mathrm{native}}\xspace}
\newcommand{\xim}{\ensuremath{\xi_\mathrm{micro}}\xspace}

\newcommand{\Tmeas}{\ensuremath{T_\mathrm{m}}\xspace}
\newcommand{\GR}{\ensuremath{G_\mathrm{R}}\xspace}
\newcommand{\Nc}{\ensuremath{{N_\mathrm{c}}}\xspace}
\usepackage{bm}
\usepackage{array}
\usepackage{xcolor}
\usepackage{float}
\usepackage [T1]{fontenc}
\usepackage [utf8]{inputenc}
\usepackage[colorlinks]{hyperref}

\usepackage[T1]{fontenc}
\usepackage[utf8]{inputenc}
\usepackage{graphicx}
\usepackage{amsmath}
\usepackage{amssymb}
\usepackage{color}
\usepackage{xspace}
\usepackage{multirow}
\usepackage{hyperref}
\usepackage{dcolumn}
\makeatletter
\newcommand*{\balancecolsandclearpage}{%
  \close@column@grid
  \cleardoublepage
  \twocolumngrid
}
\makeatother
\graphicspath{{./figures/}{./}}
\begin{document}
\title{Quantifying memory in spin glasses}

\author{I.~Paga} \email{ilaria.paga@gmail.com}\affiliation{Department of Computing Sciences, Bocconi University, 20136 Milano, Italy
}

\author{J.~He} \email{This author performed the experiments reported in this work.}\affiliation{Department of Mechanical Engineering, The University of Texas at Austin, Austin, Texas 78712, USA}

\author{M.~Baity-Jesi}\affiliation{Eawag,  Überlandstrasse 133, CH-8600 Dübendorf, Switzerland}

\author{E.~Calore}\affiliation{Dipartimento di Fisica e Scienze della Terra, Università di Ferrara e INFN, Sezione di Ferrara, I-44122 Ferrara, Italy}

\author{A.~Cruz}\affiliation{Departamento de Física Teórica, Universidad de Zaragoza, 50009 Zaragoza, Spain}\affiliation{Instituto de Biocomputación y Física de Sistemas Complejos (BIFI), 50018 Zaragoza, Spain}

\author{L.A.~Fernandez}\affiliation{Departamento de Física Teórica, Universidad Complutense, 28040 Madrid, Spain}

\author{J.M.~Gil-Narvion}\affiliation{Instituto de Biocomputación y Física de Sistemas Complejos (BIFI), 50018 Zaragoza, Spain}

\author{I.~Gonzalez-Adalid Pemartin}\affiliation{Departamento de Física Teórica, Universidad Complutense, 28040 Madrid, Spain}

\author{A.~Gordillo-Guerrero}\affiliation{Departamento de Ingeniería Eléctrica, Electrónica y Automática, U. de Extremadura, 10003, Cáceres, Spain}\affiliation{Instituto de Computación Científica Avanzada (ICCAEx), Universidad de Extremadura, 06006 Badajoz, Spain}

\author{D.~I\~niguez}\affiliation{Instituto de Biocomputación y Física de Sistemas Complejos (BIFI), 50018 Zaragoza, Spain}\affiliation{Fundación ARAID, Diputación General de Aragón, Zaragoza, Spain}\affiliation{Departamento de Física Teórica, Universidad de Zaragoza, 50009 Zaragoza, Spain}

\author{A.~Maiorano}\affiliation{Dipartimento di Biotecnologie, Chimica e Farmacia, Università degli studi di Siena, 53100, Siena Italy}\affiliation{Instituto de Biocomputación y Física de Sistemas Complejos (BIFI), 50018 Zaragoza, Spain}

\author{E.~Marinari}\affiliation{Dipartimento di Fisica, Sapienza Università di Roma, and CNR-Nanotec, Rome Unit, and INFN, Sezione di Roma1, 00185 Rome, Italy}

\author{V.~Martin-Mayor}\affiliation{Departamento de Física Teórica, Universidad Complutense, 28040 Madrid, Spain}
 
\author{J.~Moreno-Gordo}\affiliation{Instituto de Biocomputación y Física de Sistemas Complejos (BIFI), 50018 Zaragoza, Spain}\affiliation{Departamento de Física Teórica, Universidad de Zaragoza, 50009 Zaragoza, Spain}\affiliation{Departamento de Física, Universidad de Extremadura, 06006 Badajoz, Spain}\affiliation{Instituto de Computación Científica Avanzada (ICCAEx), Universidad de Extremadura, 06006 Badajoz, Spain}

\author{A.~Mu\~noz Sudupe}\affiliation{Departamento de Física Teórica, Universidad Complutense, 28040 Madrid, Spain}
  
\author{D.~Navarro}\affiliation{Departamento de Ingeniería, Electrónica y Comunicaciones and I3A, U. de Zaragoza, 50018 Zaragoza, Spain}

\author{R.L.~Orbach}
\affiliation{Texas Materials Institute, The University of Texas at Austin, Austin, Texas  78712, USA}

\author{G.~Parisi}\affiliation{Dipartimento di Fisica, Sapienza Università di Roma, and CNR-Nanotec, Rome Unit, and INFN, Sezione di Roma1, 00185 Rome, Italy}

\author{S.~Perez-Gaviro}\affiliation{Departamento de Física Teórica, Universidad de Zaragoza, 50009 Zaragoza, Spain}\affiliation{Instituto de Biocomputación y Física de Sistemas Complejos (BIFI), 50018 Zaragoza, Spain}
  
\author{F.~Ricci-Tersenghi}\affiliation{Dipartimento di Fisica, Sapienza Università di Roma, and CNR-Nanotec, Rome Unit, and INFN, Sezione di Roma1, 00185 Rome, Italy}

\author{J.J.~Ruiz-Lorenzo}\affiliation{Departamento de Física, Universidad de Extremadura, 06006 Badajoz, Spain}\affiliation{Instituto de Computación Científica Avanzada (ICCAEx), Universidad de Extremadura, 06006 Badajoz, Spain}

\author{S.F.~Schifano}\affiliation{Dipartimento di Scienze dell'Ambiente e della Prevenzione Università di Ferrara and INFN Sezione di Ferrara, I-44122 Ferrara, Italy}

\author{D.L.~Schlagel} \affiliation{Division of Materials Science and Engineering, Ames Laboratory, Ames, Iowa 50011, USA}

\author{B.~Seoane}\affiliation{Université Paris-Saclay, CNRS, INRIA Tau team, LISN, 91190 Gif-sur-Yvette, France}\affiliation{Instituto de Biocomputación y Física de Sistemas Complejos (BIFI), 50018 Zaragoza, Spain}

\author{A.~Tarancon}\affiliation{Departamento de Física Teórica, Universidad de Zaragoza, 50009 Zaragoza, Spain}\affiliation{Instituto de Biocomputación y Física de Sistemas Complejos (BIFI), 50018 Zaragoza, Spain}

\author{D.~Yllanes}\affiliation{Fundaci\'on ARAID, Diputaci\'on General de
  Arag\'on, Zaragoza, Spain}\affiliation{Instituto de Biocomputación y Física de Sistemas Complejos (BIFI), 50018 Zaragoza, Spain}
\affiliation{Zaragoza Scientific Center for Advanced Modeling (ZCAM), 50018 Zaragoza, Spain}
\affiliation{Chan Zuckerberg Biohub, San Francisco, CA, 94158}

\collaboration{Janus Collaboration}

\date{\today}
%%%%%%%%%%%%%%%%%%%%%%%%%%%%%%%%%%%%%%%%%%%%%%%%%%%%%%%%%%%%%%%%%%%%%%
\begin{abstract}
\noindent 
Rejuvenation and memory, long considered the distinguishing features
of spin glasses, have recently been proven to result from the growth
of multiple length scales. This insight, enabled by simulations on the
Janus~II supercomputer, has opened the door to a quantitative
analysis.  We combine numerical simulations with comparable
experiments to introduce two coefficients that quantify memory. A third coefficient has been recently presented by
Freedberg et al.  We show that these coefficients are physically
equivalent by studying their temperature and waiting-time dependence.
\end{abstract}

\maketitle 
Memory is among the most striking features of far-from-equilibrium systems~\cite{keim:19}, including granular materials~\cite{bandi:18}, phase separation in the early universe~\cite{koide:06} and, particularly, glass formers~\cite{ozon:03,bellon:00,yardimci:03,bouchaud:01b,mueller:04,scalliet:19,pashine:19}.
Whether a universal mechanism is responsible for memory in all these materials 
is unknown, but spin glasses stand out~\cite{lefloch:92,jonason:98,lundgren:83,jonsson:00,hammann:00,vincent:07,vincent:23}. On the one hand, memory effects are particularly strong in these systems ---perhaps because of the large attainable coherence lengths~\cite{zhai:19,zhai-janus:20a,zhai-janus:21}. More importantly, their 
dynamics is now understood in great detail.
Indeed, to model protocols where temperature is varied, one must first understand
the nonequilibrium evolution at constant temperature. In other words, before
tackling memory, rejuvenation and aging should be mastered. These intermediate
steps, including the crucial role of temperature chaos, have now been taken for spin glasses~\cite{janus:18,janus:21,zhai:22}.

In the context of spin glasses, rejuvenation is the observation that when the system is aged at a temperature $T_1$ for a time \twone, and then cooled to a sufficiently lower $T_2$, the spin glass reverts apparently to the same state it would have achieved had it been cooled directly to $T_2$.  That is, its apparent state is independent of its having approached equilibrium at temperature $T_1$. However, when the spin glass is then warmed back to temperature $T_1$, it appears to return to its aged state, hence memory.

Conventional wisdom has long ascribed rejuvenation to temperature chaos ---the notion that
equilibrium states at close temperatures are unrelated--- and memory
was experimentally exhibited long ago~\cite{lefloch:92,jonason:98} but these effects remained
unassailable to numerical simulations in three-dimensional systems~\cite{komori:00,picco:01,berthier:02,takayama:02,maiorano:05,jimenez:05}. 
This state of affairs has recently changed~\cite{janus:23}, thanks to the combination of multiple advances: the availability of the Janus~II supercomputer~\cite{janus:14} and of single-crystal experiments~\cite{zhai:19}, accessing much larger coherence lengths; the establishment of a relation between
experimental and numerical time scales~\cite{zhai-janus:20a, zhai-janus:21}; and the quantitative modelling of nonequilibrium temperature chaos~\cite{janus:21}.
Ref.~\cite{janus:23} has not only demonstrated memory numerically  and related rejuvenation to temperature chaos, but also shown that both effects are ruled by multiple length scales, setting the stage for 
a more quantitative study. 

Here we introduce two coefficients to quantify memory, one experimentally accessible and the other adapted to numerical work. The example of temperature chaos 
has shown that such indices are key to a comprehensive theory~\cite{fernandez:13,fernandez:16,billoire:18}. In principle, the only constraints
for such a coefficient $\mathcal C$ are that $\mathcal C=1$ means perfect memory 
and $\mathcal C=0$ means that the memory has been totally erased. Many choices could satisfy these conditions: besides the coefficients introduced herein, \cite{freedberg:23} presents an alternative based on a different observable. Fortunately,
the length scales discussed in~\cite{janus:23} allow us to express these coefficients as
smooth functions of similar scaling parameters and to demonstrate that they have the same physics as temperature and waiting time are varied. 

\paragraph*{Protocols.}
 In both experiment and simulations~\cite{orbach-janus:23}, we have a three-step procedure: (i) the system is quenched  to an aging temperature $T_1 < \Tg$ (\Tg is the glass temperature) and relaxes for a time $\twone$. In simulations, this quench is instantaneous, while in experiment, it is done at $\approx$~10\,K/min. A protocol where $T$ is kept constant after the initial quench is termed native.
  (ii) The system is then quenched to $T_2<T_1$, where it evolves for time $\twtwo$. (iii) The system is raised back to $T_1$, instantaneously in simulations, and at the same rate as cooling in experiment. After a short time at $T_1$ ($2^{10}$ time steps in simulations), the dynamics are compared to the native system, which has spent $\twone$ at $T_1$. The temperature drop $T_1- T_2$ is chosen to ensure that temperature chaos (and, hence, rejuvenation) is sizeable~\cite{zhai:22,janus:23}. Our experiments are performed on a sample with $\Tg=41.6$\,K.  For experimental parameters see Table~\ref{tab:exp} (Table~\ref{tab:attemps_NUM} for simulations).

\paragraph*{Length scales.} Memory and  rejuvenation are ruled by several related length scales \footnote{See Supplemental Material [url] , which
includes Refs. \cite{janus:13, mydosh:93} for details}, of which 
only one is experimentally accessible ---$\xiZ$, related to the Zeeman effect~\cite{joh:99}.
In simulations, the basic length is the size of glassy domains in a native protocol, \ximN.

We regard rejuvenation as a consequence of temperature chaos. If the temperature drop meets the chaos requirement, namely the chaos length scale set by $T_1-T_2$ is small compared to $\ximN(T_1,\twone)$ \cite{zhai:22,janus:23}, the system that has aged at the starting temperature $T_1$ for $\twone$ ``rejuvenates''  at the cold temperature $T_2$.  That is, preexisting correlated spins are ``frozen'' dynamically at $T_2$ and a
new correlated state of size $\zeta(T_2,\twtwo)$ forms, where $\twtwo$ is the waiting time at $T_2$.  The newly created correlated state at $T_2$ is independent from that formed at $T_1$. 

As $\tw$ increases at $T_2$, the newly correlated state at $T_2$ grows larger, to a maximum size set by the final time at $T_2$, $\twtwo$.  Upon heating back to $T_1$, the two correlated states {\it interfere}, causing a memory loss that will be seen both in experiment and simulations.
\begin{figure}[t]
	\centering
	\includegraphics[width = 1\columnwidth]{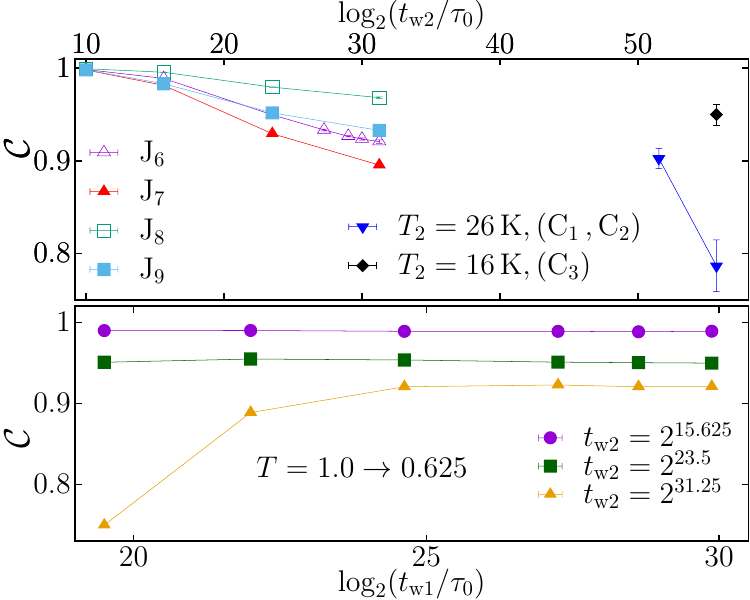}
	\caption{\textbf{Memory coefficient} as a function of \twone and \twtwo. Top:  Memory loss as a function of time \twtwo both for our
  experiments (Table~\ref{tab:exp}) and for our simulations (Table~~\ref{tab:attemps_NUM}).
   The ``natural'' timescale $\tau_0$ is $1.86\times10^{-13}$ seconds for the experimental data and $1$ lattice sweep for simulations. Bottom: Memory coefficient for fixed $t_\mathrm{w1}$ [fixed $\xi_\mathrm{micro}(T_1,\twone)$] and several \twtwo.}

\label{fig:memory_vs_time}
\end{figure}

\begin{table}
\begin{ruledtabular}
\begin{tabular}{c c c c c c c}
Protocol&$T_1$\ (K)&$\twone$\ (h)&$T_2$\ (K)&$\twtwo$\ (h)&\Tmeas(K)&$\xi/a$\\\hline
N$_1$&30&1  &---&---&30&13.075\\
N$_2$&26&1/6&---&---&26&8.1011\\
N$_3$&26&3  &---&---&26&11.961\\
N$_4$&16&3  &---&---&16&6.3271\\
R$_1$&30&1  &26 &3  &26&11.787\\
  C$_1$&30&1  &26 &1/6&30 &12.621\\
C$_2$&30&1  &26 &3  &30 &12.235\\
C$_3$&30&1  &16 &3  &30 &12.846\\
\end{tabular}
\end{ruledtabular}
\caption{{\bf Parameters of our experiments.} We use the abbreviations N (native), R (rejuvenation) and C (cycle), see  paragraph \emph{Protocols} for discussion. Unlike in rejuvenation protocols, in temperature cycles the temperature is brought back to $T_1$ before measurements start. We indicate the measuring temperature $\Tmeas$, and ---whenever possible--- the value of $\xi$, in units of the average distance between nearest-neighbour spins. Measurements start after $\twone$ for native protocols ($\twone+\twtwo$ for cycle or rejuvenation protocols).  The data for $\text N_2$ were taken at small values of $H$ because non-linear effects enter for larger values of $H$ at small \tw.}\label{tab:exp}
\end{table}

Simulations can easily access $\ximN$ and $\zeta$ ~\cite{janus:08b,janus:09b,janus:18,janus:23}.  Experiments measure
instead the number of correlated spins $N_\text{c}$~\cite{joh:99}. Equivalently, in simulations, the native protocol results in
\begin{equation}\label{eq:Nc}
\Nc^\text{native}(T,\tw)= \bigl[\ximN(T,\tw)\bigr]^{D-\theta/2}\, ,  
\end{equation}
where $\theta$ is the replicon exponent~\cite{janus:17,zhai-janus:20a,zhai-janus:21,orbach-janus:23}. If the chaos condition is met, and only in this case,
the jump protocol results in, concomitantly, $\Nc^{\text{jump}}(T_2,\twtwo) \propto [\zeta(T_2,\twtwo)]^{D-\theta/2}.$
Therefore, when chaos is strong enough, we have a dictionary relating the experimentally accessible $\Nc$ to these two basic length scales~\cite{janus:23} (see also SM).

\paragraph*{Qualitative behavior of the memory coefficient.} According to 
the above, one would expect memory to depend on the 
ratio $\zeta(T_2,\twtwo)/\ximN(T_1,\twone)$: the smaller the ratio, the larger
the memory. Because both lengths are increasing functions of time, then (i) memory should increase with increasing \twone, everything else held constant.  Concomitantly, (ii) memory should decrease with increasing \twtwo, everything else held constant. Finally, (iii) if $\Delta T = T_1 - T_2$ increases, everything else held constant, $\zeta(T_2,\twtwo)$ will progressively decrease, and memory should increase. The memory coefficients that we define below, plotted in Fig.~\ref{fig:memory_vs_time}, display precisely these predicted variations. The experimental data in Fig.~\ref{fig:memory_vs_time} agree with this expectation,
which we shall test in our simulations through a scaling analysis.

\begin{figure}
    \centering
	\includegraphics[width = 1\columnwidth]{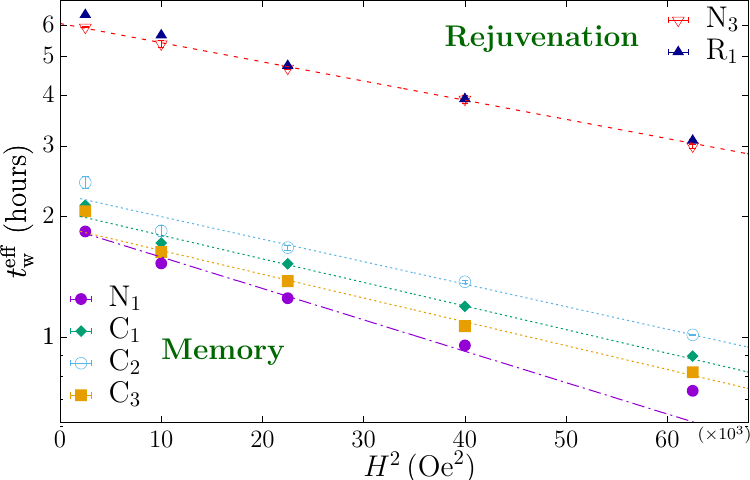}
    \caption{\textbf{Rejuvenation}  (top) and \textbf{memory} ({bottom}), as studied through the Zeeman-effect method of Ref.~\cite{joh:99} on a single crystal of CuMn. Our thermal protocols are described in Tab.~\ref{tab:exp}. We show 
    $\teffw$ vs $H^2$ in logarithmic scale [$\teffw(H)$ is the time at which  
    $S(t,H)=H^{-1}\mathrm{d}M/\mathrm{d}\log t$ peaks]. The number of correlated spins $N_{\text{c}}$ for a given protocol is proportional to the slope of the corresponding curve. The values of $\teffw$ essentially coincide for native protocol N$_3$ and for rejuvenation protocol R$_1$ at $T_2=26$\,K (the previous relaxation at $T_1=30$\,K undergone in protocol R$_1$ turns out to be  irrelevant at 26\,K). As for the memory coefficient, see Eq.~\eqref{eq:C-Zeeman}, the
    slope of native protocol N$_1$ is approached either by decreasing the relaxation time $\twtwo$ at fixed $T_2$ (compare cycles C$_1$ and C$_2$) or by lowering $T_2$ at fixed $\twtwo$ (compare cycles C$_2$ and  C$_3$).}
    \label{fig:exp}
\end{figure}

\paragraph*{Experimental definition of the memory coefficient.} We define
a memory coefficient from the number of correlated spins $N_{\text{c}}$, see Eq.~\eqref{eq:Nc}.  We shall first show that
rejuvenation can be observed from $N_{\text{c}}$, thus confirming that temperature chaos is strong enough given our choice of temperatures and waiting times. 

A small sample was cut from a single crystal of CuMn, 7.92 at.\%, with $T_\mathrm{g} = 41.6$\,K. $T_1$ and $\twone$ in Table~\ref{tab:exp} were chosen to facilitate direct comparison with the
results of Freedberg et al.~\cite{freedberg:23}, who defined another memory coefficient from the linear magnetic susceptibility. The lower temperature, $T_2$, was variable, as was the waiting time $\twtwo$. Our measurements of $N_{\text{c}}$ follows Ref.~\cite{joh:99} and are illustrated in Fig.~\ref{fig:exp}--top.
For the reader's convenience, we briefly recall the main steps leading to the measurement of $N_{\text{c}}$ (see Supplemental Material [url] for further details).

After the sample has undergone the appropriate preparatory protocol, we probe its dynamic state by switching on a magnetic field. We set time $t=0$ when the field is switched on and record the magnetization $M(t)$, which grows steadily from $M(t=0)=0$, to obtain the dynamic response function  $S(t,H)=H^{-1}\mathrm{d}M/\mathrm{d}\log t$.
For native protocols and $H\rightarrow 0$, the peak of $S(t)$ against $t$ occurs approximately at the effective time
$\teffw\approx\twone$.  As $H$
increases,  the peak moves to shorter $\teffw(H)$.
The slope of the plot of $\log \teff~\mathrm{vs}~H^2$ equals the product of \Nc times the field-cooled susceptibility per spin:
($\chi_{\text {FC}}/{\text {spin}})\Nc$ ---$\chi_{\text {FC}}$ is roughly constant over the measured temperature range. We thus gain access to \Nc.

We compare in Fig.~\ref{fig:exp}--top the outcome of the above measuring procedure as obtained in protocols N$_3$ and  R$_1$ (see Table~\ref{tab:exp}). The two systems evolve for the same time at 26\,K. The
values obtained for $\log \teff$ turned out to be nearly identical: relaxation at 30\,K does not leave
measurable traces at 26\,K, hence rejuvenation.

To quantify memory,  we  compare \Nc for systems that have undergone the temperature-cycling protocols in Table~\ref{tab:exp} with their native counterparts. These measurements are illustrated in Fig.~\ref{fig:exp}--bottom. We define the Zeeman-effect
memory coefficient as
\begin{equation}\label{eq:C-Zeeman}
{\cal C}_{\text{Zeeman}}={N_{\text{c}}^{\text{cycle}}}/{N_{\text{c}}^{\text{native}}}\,.
\end{equation}
Note that in both cases measurements are carried out at $T_1$ (the difference lies in
the previous thermal history of the sample), hence the ratio in Eq.~\eqref{eq:C-Zeeman} is just the ratio of the corresponding slopes in Fig.~\ref{fig:exp}--bottom. Our results for ${\cal C}_{\text{Zeeman}}$ are given in Fig.~\ref{fig:memory_vs_time}.
\begin{table}[t] 
\begin{ruledtabular}
\begin{tabular}{c c c c c}
             $T$-{drop} & $T_1$ & $T_2$ & $t_\mathrm{w1}$ & $\xi_\mathrm{micro}(T_1,t_\mathrm{w1})$\\ [1pt]
                        \hline  \\ [-1.9ex]
                        J$_1$ & 1.0 & 0.625   & $2^{19.5}$  & 8.038(1)\\
                        J$_2$ & 1.0 & 0.625   & $2^{22}$  & 10.085(15)\\
                        J$_3$ & 1.0 & 0.625   & $2^{24.625}$  & 12.75(3)\\
                        J$_4$ & 1.0 & 0.625   & $2^{27.25}$  & 16.04(3)\\
                        J$_5$ & 1.0 & 0.625   & $2^{28.625}$  & 18.08(5)\\
                        J$_6$ & 1.0 & 0.625   & $2^{29.875}$  & 20.20(8)\\
                        J$_7$ & 1.0 & 0.7   & $2^{29.875}$  & 20.20(8)\\
                        J$_8$ & 0.9 & 0.5   & $2^{31.25}$  & 16.63(5)\\ 
                        J$_9$ & 0.9 & 0.7   & $2^{31.25}$  & 16.63(5)\\ [0pt] 
\end{tabular}
\end{ruledtabular}
\caption{{\bf  Building blocks \boldmath ($T$-drop) for temperature cycles in our simulation}. The initially disordered system relaxes for a time $\twone$ at temperature $T_1$, reaching a coherence length
$\xi_\mathrm{micro}(T_1,t_\mathrm{w1})$. A $T$-drop is given by $(T_1,T_2,\twone)$. To specify a temperature cycle one needs in addition $\twtwo$, see the \emph{Protocols} paragraph. Only J$_9$ does not meet the requirements for temperature chaos~\cite{janus:23}.}
   \label{tab:attemps_NUM}
\end{table}

\paragraph*{A memory coefficient from simulations.\/} 
We extend \cite{janus:23} to quantify memory 
through a computation  at the verge of current
capabilities~\footnote{Our simulations will closely follow~\citep{janus:23} (for the sake of completeness we describe them in Table~\ref{tab:attemps_NUM} and in the Supplemental Materials ).}. We look for memory through the quantity that can be extracted most accurately from simulations: the spin-glass correlation function at $H\!=\!0$,
$\GR({\boldsymbol r},t;p)= \overline { \langle q^{(a,b)}(\boldsymbol{x},t) q^{(a,b)}(\boldsymbol{x+r},t) \rangle_p}\,.$
Here, $q^{(a,b)}(\boldsymbol{x},t) \equiv \sigma^{(a)}(\boldsymbol{x},t)\,\sigma^{(b)}(\boldsymbol{x},t)$ where $(a,b)$ label different real replicas and $\langle \cdots \rangle_p$ stands for the thermal average after a temperature cycle built from $T$-drop $p$ (see Table~\ref{tab:attemps_NUM} and \emph{Protocols}) \footnote{Due to rotational invariance~\cite{janus:09b}, \GR essentially depends on $r=|\boldsymbol{r}|$.}.

Our starting observation is that the experimental determination of ${\cal C}$, Fig.~\ref{fig:exp}, relies on the nonlinear response to the magnetic field. Interestingly enough, the equilibrium nonlinear susceptibility is proportional to the integral of $r^2 \GR$ with $r\in(0,\infty)$. Thus, following~\cite{janus:17,janus:17b,zhai-janus:20a,zhai-janus:21}, we generalize this equilibrium relation by computing these integrals using the nonequilibrium correlation function $\GR$.

\begin{figure}[t]
 \includegraphics[width=\columnwidth]{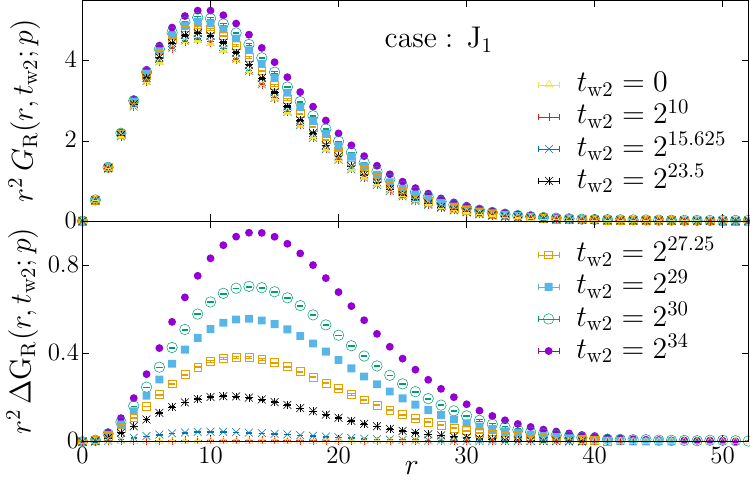}
  \caption{\textbf{Numerical construction of a memory coefficient}. {Top}: the curves $r^2 \GR(r,\twtwo;p)$ vs. $r$  show small, but detectable, differences  
  as $\twtwo$ varies, for temperature {cycles} built from the $T$-drop J$_1$ in Table~\ref{tab:attemps_NUM} ($ \twone=2^{19.5}$).  {Bottom}: as in Top, but subtracting the native \GR from the correlation function of the temperature-cycled system. As the time spent at the colder temperature $T_2$ increases, the system loses memory. The memory loss is evinced by the increasing signal in $r^2\Delta \GR$.}
  \label{fig:memory_definition_num}
\end{figure}

Fig.~\ref{fig:memory_definition_num}-top exhibits a small but detectable difference in  the behavior of $r^2 \GR$ as
$\twtwo$ varies in cycles built from $T$-drop $J_1$
in Table~\ref{tab:attemps_NUM}. This success has encouraged to consider the curve with $t_\mathrm{w2}=0$ as the reference curve~\footnote{A native protocol can be regarded as a cycle with $\twtwo\!=\!t_\mathrm{w3}\!=\!0$ ($\twone+2^{10}$ is indistinguishable from $\twone$ within our numerical accuracy for native runs).}. We evaluate effects due to $t_\mathrm{w2}$ as
$
\Delta \GR(r, t_\mathrm{w2}; p) = \GR(r,t_\mathrm{w2};p) - \GR(r, 0; p)\,.$
 $\Delta \GR$ measures the effects caused by aging at $T_2$ for a waiting time $\twtwo$.  If $T_1-T_2$ is sufficient for full chaos to develop at $T_2$, then $\Delta G_{\text {R}}(r,\twtwo;T_1)$ represents the competing correlation length that interferes with the native correlation length established at $T_1$.

 From Fig.~\ref{fig:memory_definition_num}--bottom, it is
 natural to define the numerical memory coefficient $\mathcal{C}_\mathrm{num}$ as
\begin{equation}
\label{eq:Memory_coefficient}
\mathcal{C}_\mathrm{num} = 1-  \frac{ \int_{0}^{\infty} \mathrm{d}r\, r^2 \,\Delta G_{\text{R}}(r,\twtwo;p)}{\int_{0}^{\infty} \mathrm{d}r\, r^2 \, G_{\text{R}}(r,0;p)}\,.
\end{equation}
Given the accuracy we achieve for $\Delta G_{\text{R}}(r,\twtwo;p)$, $\mathcal{C}_\mathrm{num}$ is sensitive to even tiny differences in the state of the system just before and just after the temperature cycle.
Some scaling relations linking $\mathcal{C}_\mathrm{num}$, $\mathcal{C}_\mathrm{Zeeman}$ and the coefficient proposed in Ref.~\cite{freedberg:23} are discussed in the Supplemental Material [url] which includes also  Refs.\cite{fernandez:23, nordblad:97}.
Our results for $\mathcal{C}_\mathrm{num}$ are presented in Fig.~\ref{fig:memory_vs_time}, for several temperature cycles 
built from the temperature drops in Table ~\ref{tab:attemps_NUM}.
The values of $T_1$, $T_2$ and $\twone$ were chosen to meet the chaos requirement proposed in~\cite{zhai:22,janus:23} (the only exception is $\mathrm{J}_9$, used as a testing case for the scaling analysis, below). It is comforting that, even when
the ratio $\twtwo/\twone$ is as large as in Fig.~\ref{fig:memory_definition_num}, we still obtain $\mathcal{C}_\mathrm{num} > 0.75$.

\paragraph*{Discussion.} Memory can be quantified in several ways: we have proposed two
such memory coefficients, $\mathcal{C}_{\text{Zeeman}}$ and $\mathcal{C}_{\text{num}}$, respectively, adapted to experimental and numerical computation. Each coefficient is used in a different time scale, see  Fig.~\ref{fig:memory_vs_time}.
Furthermore,~\cite{freedberg:23} proposes yet another experimental coefficient, $\mathcal{C}_{\chi''}$,
based on the linear response to a magnetic field (rather than the nonlinear responses considered herein). It is obvious that more options exist. Hence, it is natural to ask what (if any) is the relationship between these coefficients.

We look for this relationship in the two length scales that rule our nonequilibrium dynamics, namely ${\zeta(T_2,\twtwo)}$ and  ${\ximN(T_1,\twone)}$. If we succeed in expressing
our coefficients as simple functions of these two lengths, we shall naturally link different memory definitions. 

\begin{figure}
	\centering
	\includegraphics[width = 1\columnwidth]{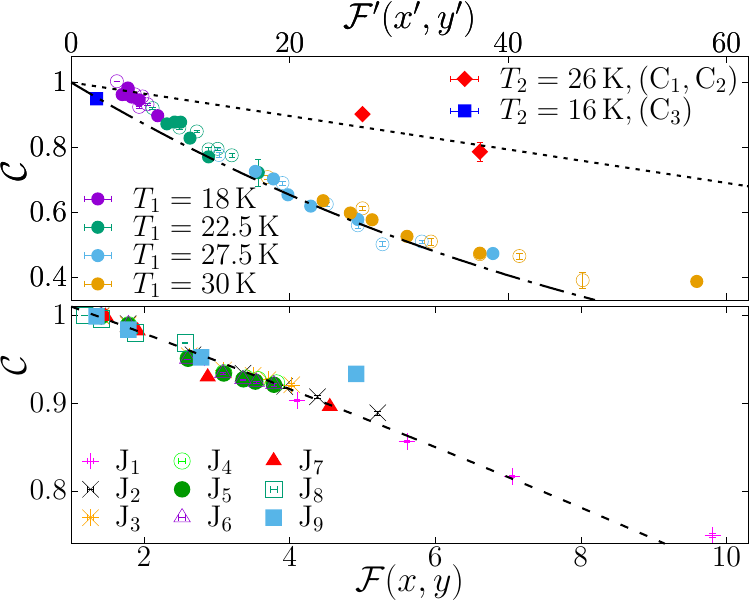}
	\caption{{Top:} Experimental memory coefficients $\mathcal{C}_{\chi''}$
(circles, data from ~\cite{freedberg:23}; $T_1=T_2+4$\,K) and
$\mathcal{C}_\mathrm{Zeeman}$ (rhombus and squares, see
Table~\ref{tab:exp}) versus the scaling function
in Eq.~\eqref{eq:scaling_law_gen}. See text for the
computation of the $x'$ and $y'$ proxies.
Empty circles refer to $\twone=1 \, \mathrm{h}$ (full circles for
$\twtwo= 3 \, \mathrm{h})$. The dotted line is a fit to a straight
$\mathcal{C}_\mathrm{Zeeman}=1-\alpha \mathcal{F}'$.  The dashed line is $(1-\alpha \mathcal{F}')^{3.9}$ and represents the experimental findings for $\mathcal{C}_{\chi''}$ surprisingly well.  {Bottom:} As in Top, for memory coefficient $\mathcal{C}_\mathrm{num}$ [$x$ and $y$ are given in Eq.~\eqref{eq:def-x-and-y}], see Table~\ref{tab:attemps_NUM} for details.}
	\label{fig:scaling_law}
 \end{figure}

Specifically, we consider two variables $x$ and $y$:
\begin{equation}\label{eq:def-x-and-y}
x\!=\! \Big[\frac{\zeta(T_2,\twtwo)}{\ximN(T_1,\twone)}\Big]^{D-\theta/2}\!\!,\
y=\frac{T_1}{\Tg} \, \zeta(T_2,\twtwo)\,.
\end{equation}
Both scaling variables, $x$ and $y$, are approximately accessible to experiment through  $N_{\text{c}}^{\text{jump}}$ and
$N_{\text{c}}^{\text{native}}$;
we shall name their experimental proxies $x'$ and $y'$~\footnote{We computed $x'$ and $y'$ from the approximation $\zeta(T_2,\twtwo)\approx\xim(T_2,\twtwo)$~\citep{janus:23} and the approximate law $\xim(T,t_{\mathrm{w}})/a\approx 0.58 (t/t_{\mathrm{w}})^{c_2 T/T_{\mathrm{g}}}$ with
$c_2=0.104$ and $\tau=0.186$\,ps~\protect{\cite{zhai:19}}. Indeed, our measurements of $\xi$ in Table~\ref{tab:exp} rather follow $\xi/a\approx 0.58 (t/t_{\mathrm{w}})^{c_2 T/T_{\mathrm{g}}} +3.9$.  We omit the constant  background $3.9$ to diminish corrections to scaling.}. Therefore, we
seek numerical constants $a_1$ and $a_2$, [see SM for details], such that the
$\mathcal{C}_{\mathrm{num}}$ from all our temperature cycles
fall onto a single function of
\begin{equation}\label{eq:scaling_law_gen}
    \mathcal{F}(x,y)= y \,  [1+a_1 x+a_2 x^2]\,.
\end{equation}
Setting aside $T$-drop J$_9$, which does not meet the chaos condition, the overall linear behavior in Fig.~\ref{fig:scaling_law}--bottom is reassuring.

With appropriate  $a'_1$ and $a'_2$ in Eq.~\eqref{eq:scaling_law_gen}, see SM for details, the data for
$\mathcal{C}_{\chi''}$~\cite{freedberg:23} also fall onto a smooth function of $\mathcal{F}'(x',y')$, Fig.~\ref{fig:scaling_law}--top. There is a problem, however:
$\mathcal{C}_{\chi''}$ goes to one for $\twtwo$ significantly larger than zero ($\mathcal{F'}=0$ only at
$\twtwo=0$). The same problem afflicts $\mathcal{C}_{\mathrm{num}}$, albeit to a lesser degree. 
Interestingly, when plotted
as a function of $\mathcal{F}'$, $\mathcal{C}_\mathrm{Zeeman}$ is compatible with a straight line
that goes through $\mathcal{C}=1$ at $\mathcal{F}'=0$, as it should. So, at least in this respect, $\mathcal{C}_\mathrm{Zeeman}$ is the most sensible coefficient.

Perhaps more importantly, the scaling representation (dashed line in Fig.~\ref{fig:scaling_law}--top) evinces that, away from  the $\mathcal{C}\approx 1$ region, the relation $\mathcal{C}_{\chi''}\approx [\mathcal{C}_{\text{Zeeman}}]^{K}$
holds with $K\approx 3.9$. This relation makes it obvious that the
different memory coefficients carry the same physical information. Although the scaling in Eq.~\eqref{eq:scaling_law_gen} could be
feared to be accurate only near ${\cal C}\approx 1$, we have been lucky: our ansatz turns out to cover all reachable values of ${\cal C}$.

In summary, we have developed a quantitative formulation for memory in rejuvenated glasses utilizing measured and calculated coherence lengths. In combination, they account quantitatively for memory from experiment and simulations. We have confirmed that rejuvenation and temperature chaos are strongly related effects~\cite{janus:23}.  Our results are based on first principles without the need for extraneous parameters.  Our approach is easily extended to the many other glassy systems that exhibit these phenomena.

%%%%%%%%%%%%%%%%%%%%%%%%%%%%%%%%%%%%%%%%%%%%%%%%%%%%%%%%%%%%%%%%%%%%%%%%%%%

%%%%%%%%%%%%%%%%%%%%%%%%%%%%%%%%%%%%%%%%%%%%%%%%%%%%%%%%%%%%%%%%%%%%%%%%%%%
\begin{acknowledgments}
We are grateful for helpful discussions with S. Swinnea about sample characterization. We thank J. Freedberg and coauthors for sharing their data and 
letting us analyze them. This work was partially supported by 
the U.S. Department of Energy, Office of Basic Energy 
Sciences, Division of Materials Science and Engineering,
under Award No. DE-SC0013599. Crystal growth of the 
Cu0.92Mn0.08 sample was performed by Deborah L. Schlagel 
at the Materials Preparation Center, Ames National 
Laboratory, U.S. Department of Energy, and supported by the Department of Energy, Basic Energy Sciences, under 
Contract No. DE-AC02-07CH11358. We were partly funded 
as well by Grant No. PID2022-136374NB-C21, PID2022-
136374NB-C22, PID2020-112936GB-I00, PID2019- 
103939RB-I00, PGC2018-094684-B-C22 and PID2021- 
125506NA-I00, funded by Ministerio de Ciencia, 
Innovación y Universidades (Spain), Agencia Estatal de 
Investigación (AEI, Spain, 10.13039/501100011033), and 
European Regional Development Fund (ERDF, A way of 
making Europe). We were also partly funded by the DGA- 
FSE (Diputación General de Aragón—Fondo Social 
Europeo). This research has been supported by the 
European Research Council under the European Unions 
Horizon 2020 research and innovation program (Grant 
No. 694925—Lotglassy, G. Parisi) and by ICSC—Centro 
Nazionale di Ricerca in High Performance Computing, Big 
Data, and Quantum Computing funded by European Union
—NextGenerationEU, and the FIS 2021 funding scheme 
(FIS783—SMaC), and by the PRIN funding scheme 
(2022LMHTET—Complexity) both from Italian MUR. 
IGAP was supported by the Ministerio de Ciencia, Innovación y Universidades (MCIU, Spain) through FPU
Grant No. FPU18/02665. IP was supported by LazioInnova-
Regione Lazio under the program Gruppi di ricerca 2020— POR FESR Lazio 2014-2020, Project NanoProbe (Application code A0375-2020-36761)
\end{acknowledgments}

\appendix

\section{Our simulations on Janus II}
\label{appendix:details_num}
We simulate the Edwards-Anderson model in a cubic lattice with linear size $L=160$  and periodic boundary conditions. The Ising spins
$S_{\bm{x}}=\pm 1$ occupy the lattice nodes and interact with their nearest lattice-neighbors through the Hamiltonian
\begin{equation}\label{SMeq:H}
    H=-\sum_{\langle \bm{x},\bm{y}\rangle} J_{\bm{x},\bm{y}} S_{\bm{x}} S_{\bm{y}}\,,
\end{equation}
where the coupling constants  are  independent random variables that are fixed at the beginning of the simulation (we choose $J_{\bm{x},\bm{y}}=\pm 1$ with 50\% probability). A choice of
the $\{J_{\bm{x},\bm{y}}\}$ is named a sample. The model in Eq.~\eqref{SMeq:H} undergoes a phase transition at temperature $T_{\mathrm{c}}=1.1019(29)$~\cite{janus:13}, separating the paramagnetic phase (at high temperatures) from the spin-glass
phase (at low temperatures).

We study the nonequilibrium dynamics of the model with a Metropolis algorithm. The time unit is a full lattice sweep (roughly corresponding to one picosecond of physical time~\cite{mydosh:93}). The simulation is performed on the Janus II custom-built supercomputer~\cite{janus:14}

We consider 16 statistically independent samples. For each sample, we simulate $N_{\text{R}}=512$ independent replicas (\emph{i.e.}, $N_{\text{R}}$ system copies that share the couplings $\{J_{\bm{x},\bm{y}}\}$ but are otherwise statistically independent). Replicas are employed to compute correlation functions as explained in Refs.~\cite{zhai-janus:20a, zhai-janus:21, janus:23}

\section{A simple relation between coherence lengths}
\label{appendix:relation_coherence_lengths}

In this section we define the relevant length scales,  $\xi_\mathrm{micro}$, $\zeta$ and $\xi_\mathrm{Zeeman}$ and their relationships:
\begin{itemize}
    \item $\xi_\mathrm{micro}(\tw)$ is the size of the glassy domain.  It is estimated through the replicon propagator, $G_\mathrm{R}(r,\tw;p)$.
    \item $\xi_\mathrm{Zeeman}$ is obtained by counting the number of spins that react coherently to an external magnetic field, and thereby the volume of correlated spins subtended by the correlation length $\xi_\mathrm{Zeeman}$. This is the only experimentally accessible method for obtaining a correlation length directly.
    \item $\zeta(t_1,t_2)$ is the scale at which defects are correlated \footnote{See Refs.~\cite{janus:08b, janus:23} for more details.}.
    
\end{itemize}

\begin{figure}[t]
	\centering
	\includegraphics[width = 1\columnwidth]{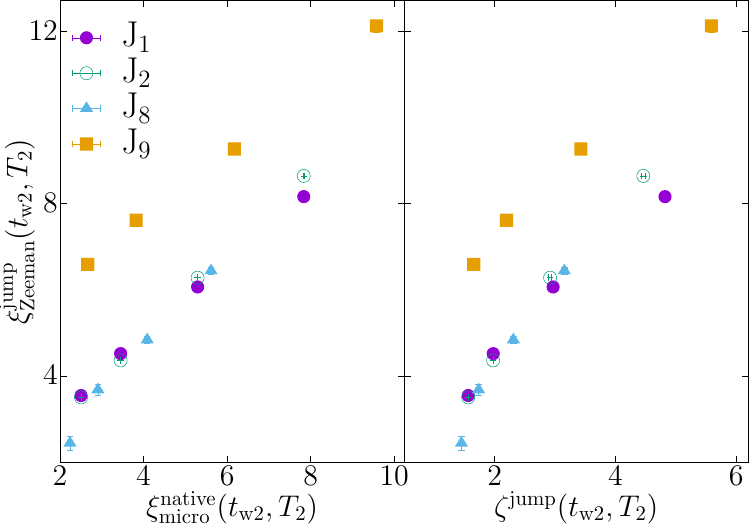}
	\caption{\textbf{Comparison between correlation lengths}. Left: we compare the behavior of the effective correlation length, $\xi_\mathrm{Zeeman}$, as a function of the microscopic correlation length, $\xi_\mathrm{micro}^\mathrm{native}(\twtwo,T_2)$ evaluated at a fixed-temperature. Right, we show the behavior of $\xi_\mathrm{Zeeman}$ as a function of the coherence scale $\zeta(\twtwo,T_2)$ calculated for varying-temperatures.  The ${\text {J}}_n$ temperature cycles  and related wait times \twone are listed in Table II in the main text.} 
	\label{fig:comparison_xiZeeman_vs_xiNat_vs_zeta}
\end{figure}

In a fixed-temperature protocol, these quantities are almost equivalent \cite{zhai-janus:20a, zhai-janus:21, janus:23}.
The scenario is more intricate in varying-temperature protocols because of  temperature chaos.
In Fig.~\ref{fig:comparison_xiZeeman_vs_xiNat_vs_zeta}, we compare these length scales in varying-temperature protocols.

As the reader can notice, if chaos is large [${\text {J}}_1$, ${\text {J}}_2$ and ${\text {J}}_8$], an equivalence exists between  $\xi_\mathrm{Zeeman}$, $\zeta$, and $\xi_\mathrm{micro}$. Otherwise [${\text {J}}_9$], the number of correlated spins is not a good proxy for $\zeta$.

\section{Scaling of \boldmath$\chi^\prime$, $\chi^{\prime \prime}$ and ${\chi}_2/V$}
\label{appendix:scaling}

In this section, we  show that the dynamical behavior of the
three susceptibilities, $\chi^\prime$, $\chi^{\prime \prime}$ and
$\hat{\chi}_2 \equiv \chi_2/V$, are similar. The former two are measured in conventional
memory experiments (\emph{e.g.}, in Ref. \cite{freedberg:23}), and the latter in numerical simulations.

We analyze first the scaling properties near the critical point. We then extend the analysis to the glass phase.

In the following we use $\epsilon \equiv (T-\Tg)/\Tg$ to denote
the reduced temperature.

On the one hand, the experimentally computed  {\em linear} magnetic susceptibilities behave as~\cite{nordblad:97}
\begin{equation}
\begin{split}
\frac{\chi_0 - \chi^\prime}{\chi_0} &\sim \epsilon^\beta G(t/ \tau)\,,\\
\chi^{\prime \prime} & \sim \epsilon^\beta H(t/\tau)\,,
\end{split}
\end{equation}
where we have taken both linear magnetic susceptibilities 
in the time domain (the relationship to frequency is a simple
Fourier transform). Above, $G(\cdot)$ and $H(\cdot)$ are two
scaling functions, and $\chi_0$ is the equilibrium value of the
linear magnetic susceptibility at the critical point.

On the other hand, the {\em nonlinear} spin-glass susceptibility 
per spin, computed in numerical simulations, scales as~\cite{nordblad:97}
\begin{equation}
{\hat \chi}_2 \sim \epsilon^{2\beta} K(t/ \tau)\,,
\end{equation}
with a suitable scaling function $K(\cdot)$. At the
critical point, there is only a critical mode and we can avoid the use of
the replicon term.

The rationale behind these scaling relations is that the overlap is
essentially the magnetization squared. The linear magnetic
susceptibilities scale with the exponent associated with the average
of the overlap $q$ (the fluctuation of the magnetization is given by
$\langle m^2\rangle$, the average of the overlap), and
$\chi_2$ is a nonlinear susceptibility per spin associated with the overlap.
Therefore, it scales with the usual exponent $\gamma=2 \beta- \nu
D$. Remember that the order parameter, the overlap, scales with the
exponent $\beta$ and that $\hat{\chi}_2=\chi_2/V$.

For the analogous scaling relations in the spin-glass phase, the out-of-equilibrium situation is dominated by the replicon mode.  The only 
diverging nonlinear susceptibility is the replicon nonlinear one, defined as
\begin{equation}
\chi_2(t) =\int_0^\infty dr\, r^2\,G_\mathrm{R}(r,t)\,.
\end{equation}

The overlap field scales with the replicon exponent $\theta$ as $q
\sim \xi^{-\theta/2}$ (see Appendix H of \cite{fernandez:23}), so we can write the dependence of these
susceptibilities on the correlation length as
\begin{equation}
\begin{split}
\frac{\chi_0 - \chi^\prime}{\chi_0} &\sim  \chi^{\prime \prime}  \sim \xi(t)^{-\theta/2}\,,\\
{\hat \chi}_2 &\sim \xi(t)^{-\theta}\,.
\end{split}
\end{equation}

We conclude from this analysis that the dynamical behaviors of these
different susceptibilities are similar. Hence, we can safely compare the results from numerical
simulations with experiments that measure $\chi^{\prime \prime}$ \cite{freedberg:23} for this reason.

\section{Experimental details and data processing}
\label{appendix:find_c2}

This section explores the experimental protocol for measuring the change in magnetization as a function of $\log t$.  In particular, it delves into the nuances of data processing for accurate identification of the time that the dynamic function $S(t)$ peaks, $\tw^{\text {eff}}(H)$. As discussed in the main body of the text, the accurate determination of $\tw^{\text {eff}}(H)$ is vital for extracting $\xi_{\text {Zeeman}}$.

Magnetization measurements were conducted using a Quantum Design MPMS system. The magnetization was gauged as the sample traversed through an array of superconducting quantum interference devices (SQUIDs). The system was set to take continuous magnetization measurements over approximately 10 hours for $\tw=1$\,h, and 30 hours for $\tw=3$\,h.

\begin{figure}[h]
	\centering
	\includegraphics[width = 1\columnwidth]{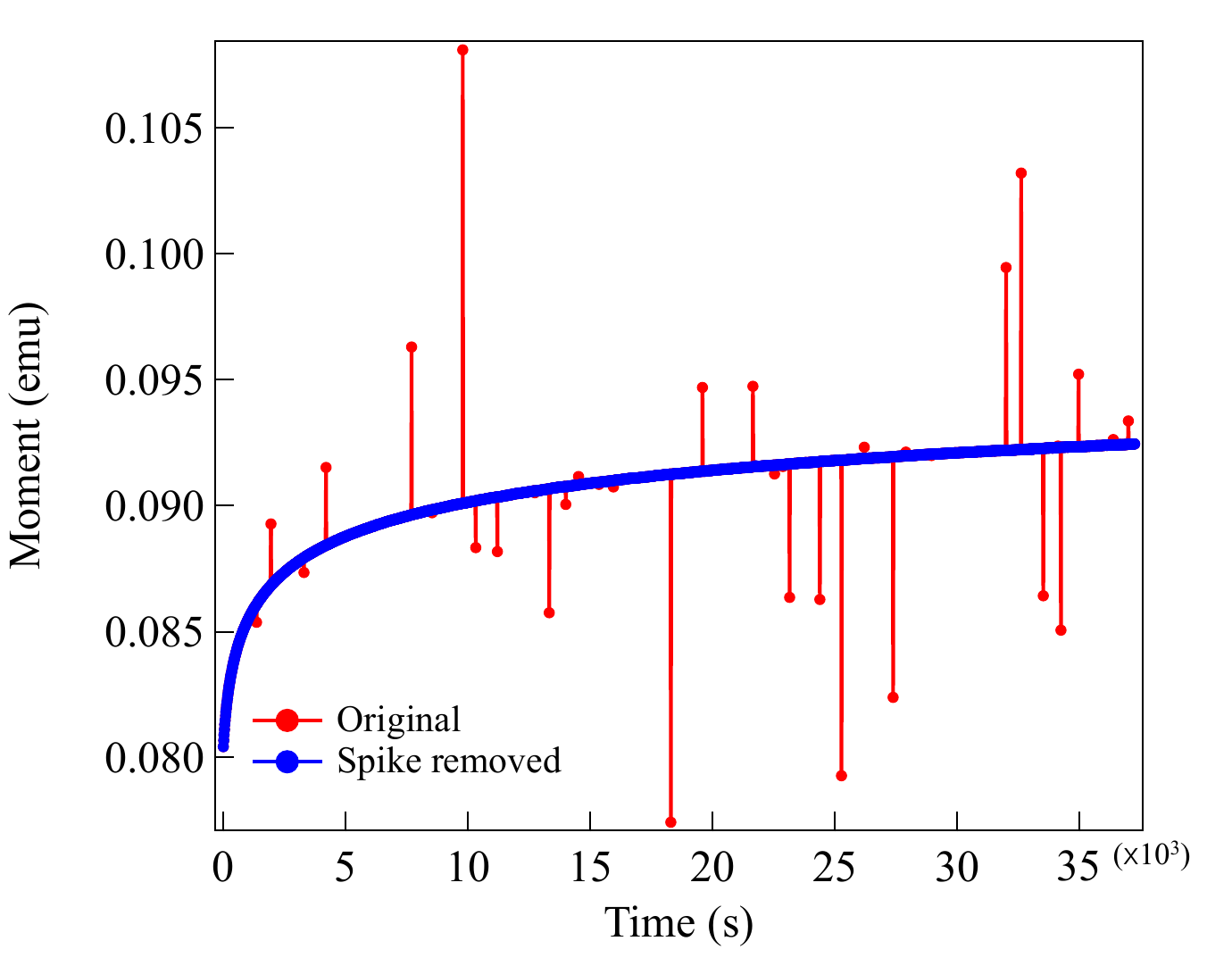}
	\caption{\textbf{Raw measurement data and the spike removed data} Data from a typical $S(t)$ measurement. (Red) The raw magnetization vs $t$ with random spikes from the SQUID measurements. (Blue) Magnetization vs $t$ with spikes removed.}
	\label{fig:C2_raw_vs_spike_removed}
\end{figure}

The magnetization as a function of time $t$ displayed intermittent spikes owing to the SQUIDs' measurements.  A representative example is exhibited in Figure \ref{fig:C2_raw_vs_spike_removed}. These aberrations, typically because of external interference with the SQUID coil, were subsequently removed in our analysis. The first derivative of the magnetization as a function of $\log t$ was then calculated.

The derivative of the raw data tends to be noisy, complicating the task of identifying the position of the peak in $S(t)$.  It was necessary to suitably smooth the $S(t)$ curve. We utilized a Chebyshev polynomial fit for the $M$ vs $t$ curve prior to computing its derivative. 

\begin{figure}[h]
	\centering
	\includegraphics[width = 1\columnwidth]{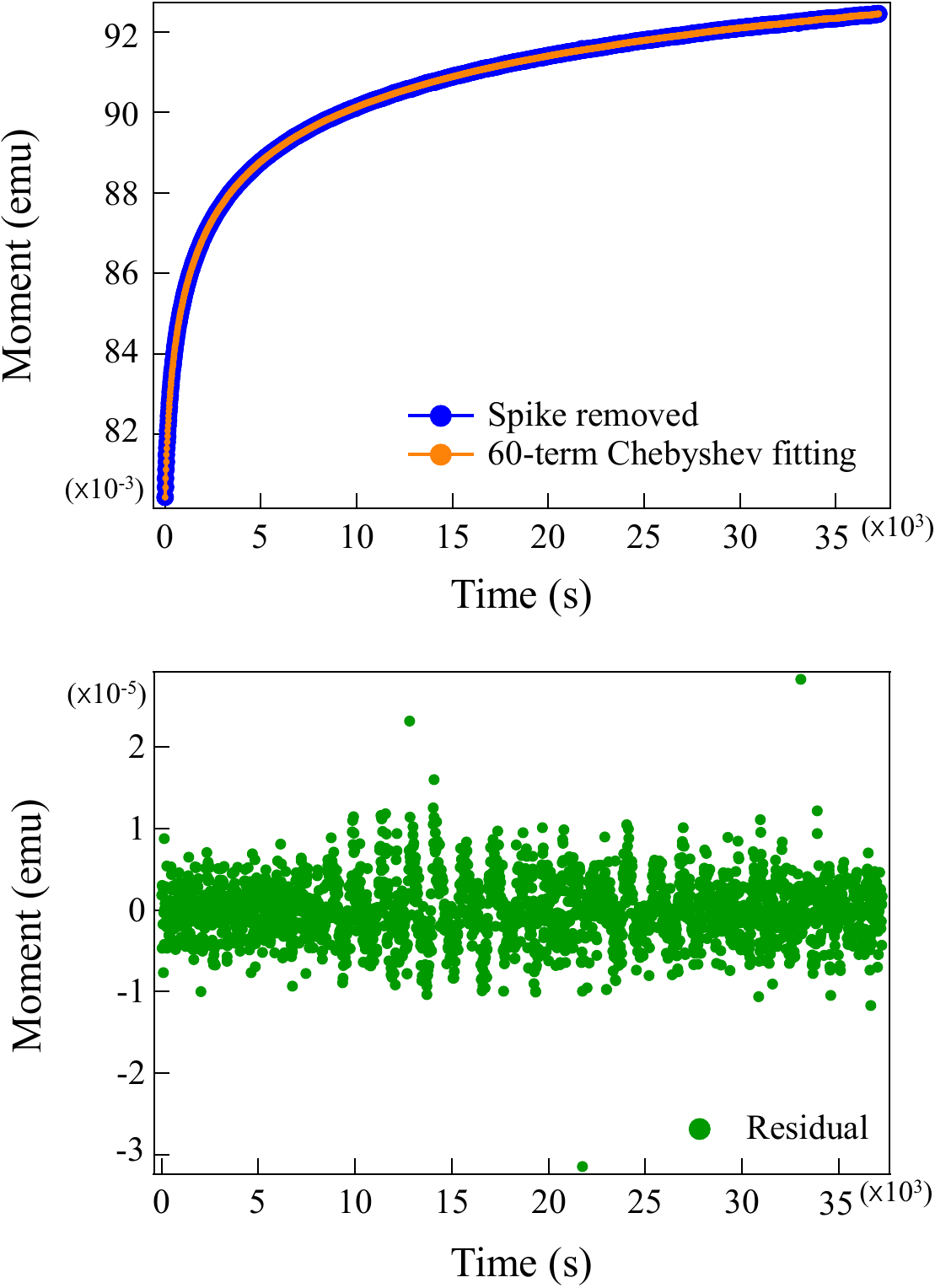}
	\caption{\textbf{Chebyshev polynomial fitting of the data and the residual} (Top) The spike removed $M$ vs $t$ data with 60-term Chebsyhev fit of the data marked in orange. (Bottom) The residual of the Chebyshev fit.}
	\label{fig:Chebshev_fitting_residual}
\end{figure}

With an appropriate number of terms, the Chebyshev polynomial fit accurately represents the raw $M$ vs $t$ data, effectively eliminating the spikes in the ${\mathrm d} M/{\mathrm d}\log t$ data produced by measurement artifacts, as illustrated in Figure \ref{fig:Chebshev_fitting_residual}. A 60-term Chebyshev polynomial fit typically depicts the $S(t)$ curve satisfactorily, with less than a 1\% residual. Using higher terms further reduces this residual. The Chebyshev polynomial fit excels over a simple box smoothing of raw data because it preserves the spline of the $M$ vs $t$ curve while simultaneously decreasing the noise in the derivative. 

\begin{figure}[!]
	\centering
	\includegraphics[width = 1\columnwidth]{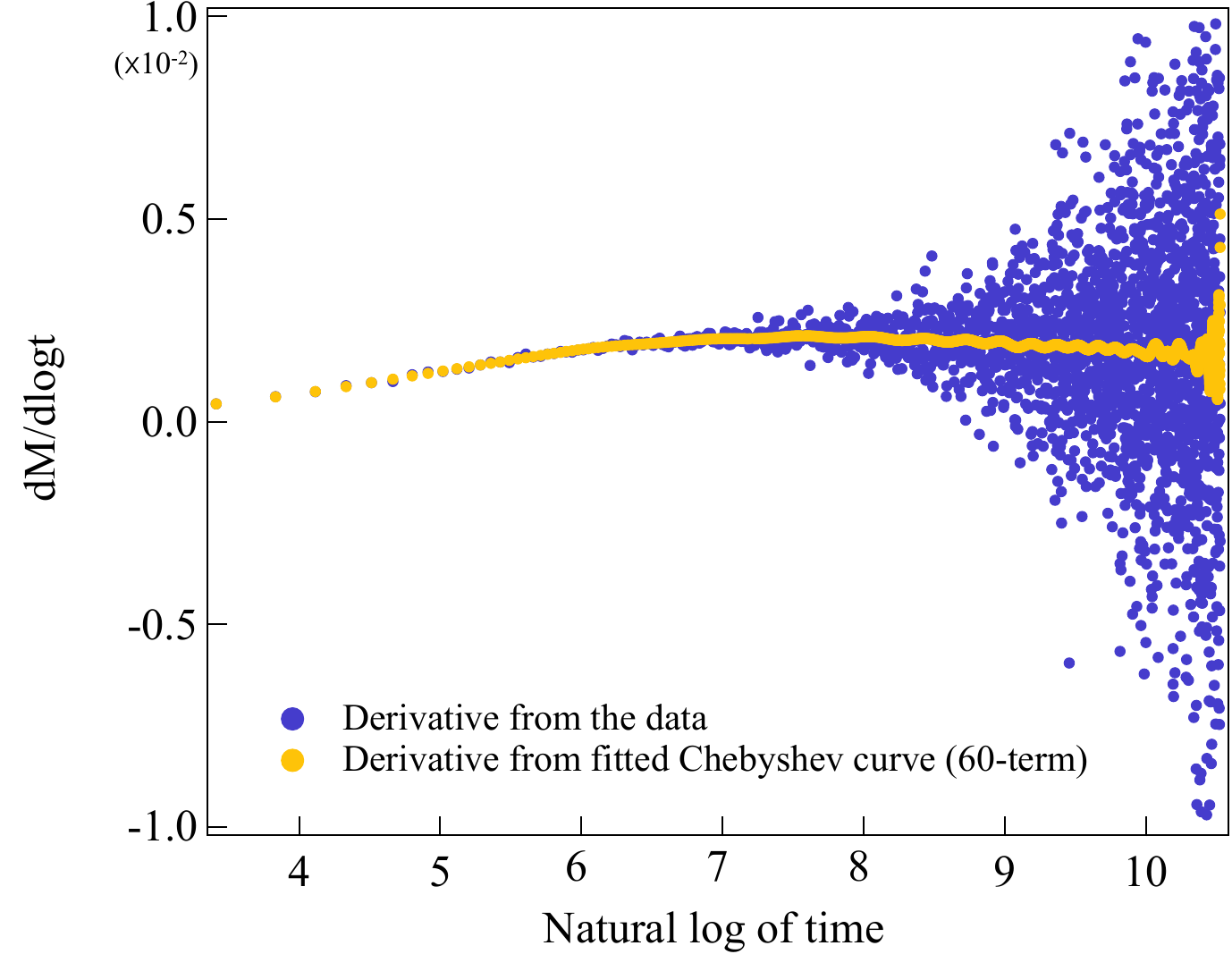}
	\caption{\textbf{Comparison of the derivative of the raw data vs the derivative of one Chebyshev fitted curve.} (Blue) The ${\mathrm d}M/{\mathrm d}\log t$ plot of raw magnetization vs $\log t$. (Yellow) The ${\mathrm d}M/{\mathrm d}\log t$ plot of a 60-term Chebyshev fit of the magnetization vs $\log t$.} 
	\label{fig:derivative_of_chebshev}
\end{figure}

Upon obtaining the derivative of the Chebyshev polynomial, we constructed a single Lorentzian peak function with a constant background:
\begin{equation}
\begin{split}
f(x) = \frac{A\gamma^2}{(x-x_0)^2+\gamma^2} + C~~.
\end{split}
\end{equation}
In this equation, $A$ denotes the amplitude of the peak, $x_0$ represents the center, and $\gamma$ indicates the peak width. As depicted in Figure \ref{fig:derivative_of_chebshev}, the derivative of the Chebyshev curve provides a relatively uncluttered peak with a fit to a Lorentzian shape.  With a sufficient number of Chebyshev polynomial terms, the fitted curve can trace the raw data precisely.

To ensure that the Chebyshev polynomial accurately represents the $M$ vs $t$ data, we iteratively conducted the fitting and peak searching procedure over a number range of terms, typically from 30 to 1200. Each iteration produced a value for the time at which the Lorentzian peaks.  We then calculated the average of these values to determine the peak time for a given $S(t)$ measurement. The error bar for the time of the peak was determined from the standard deviation of the peak times. In cases where overfitting or underfitting was apparent within certain ranges of the number of Chebyshev polynomial terms, as evinced by deviant peak values, the peak values obtained in these ranges were disregarded.

\section{Determination of the coefficients for the scaling function \boldmath$\mathcal{F}(x,y)$}\label{appendix:fit-Fxy}

As we have explained in the main text, we introduce a simple scaling function $\mathcal{F}(x,y)$ for describing both the experimental and numerical data.
Here we report the details for the determination of the constants $a_1$ and $a_2$ [and analogously  $a_1'$ and $a_2'$].\\
Our procedure consists of two steps:
\begin{itemize}
    \item we fit the data for the memory coefficient under consideration ---$\mathcal{C}_\text{num}$  or $\mathcal{C}_{\chi''}$--- to a smooth, generic function of the scaling function $\mathcal{F}(x,y)$ introduced in the main text\footnote{We employ $\mathcal{F}(x,y)$ for $\mathcal{C}_\text{num}$ and $\mathcal{F}'(x',y')$ for $\mathcal{C}_{\chi''}$.}:
    \begin{equation}
        \label{eq:memory_function}
        \mathcal{C}(x,y)= B_0-B_1 \,\mathcal{F}(x,y)-B_2 [\mathcal{F}(x,y)]^2,
    \end{equation}
    Note that there are five fit parameters,  namely $B_0$, $B_1$ and $B_2$ and the two parameters $a_1$ and $a_2$ that determine $\mathcal{F}(x,y)$.
    \item The free parameters $B_0,B_1,B_2$ are disregarded in the analysis. Instead, we keep $a_1,a_2$ (or $a_1'$ and $a_2'$ for $\mathcal{C}_{\chi''}$), to describe our data as a function of the scaling function $\mathcal{F}(x,y)$ [$\mathcal{F}'(x',y')$ in the case of $\mathcal{C}_{\chi''}$].
\end{itemize}
For the numerical data, $B_0 \simeq 1$ and $B_2=0$. The data for $\mathcal{C}_{\chi''}$, taken from Ref.~\cite{freedberg:23}, require $B_2 \neq 0$.

\bibliographystyle{apsrev4-1}
\bibliography{biblio}

\end{document}